\documentclass[useAMS,usenatbib]{mn2e}
\usepackage{graphicx}

\title[Fast photometric observations of V2069 Cyg]{\LARGE{Very fast photometric and X-ray observations of the intermediate polar \\ V2069 Cygni (RX J2123.7+4217)}}

\author[Nasiroglu et al.]{I. Nasiroglu,$^{1,2}$\thanks{E-mail: nasiroglu@mpe.mpg.de (IN); aga@astro.ia.uz.zgora.pl (AS); gok@mpe.mpg.de (GK); fwh@mpe.mpg.de (FH)} A. S\l{}owikowska,$^{3\star}$ G. Kanbach,$^{1\star}$ and F. Haberl$^{1\star}$ \\
$^{1}$Max Planck Institute for Extraterrestrial Physics, Giessenbachstrasse, 85748 Garching, Germany\\
$^{2}$University of Cukurova, Department of Physics, 01330 Adana, Turkey\\
$^{3}$Institute of Astronomy, University of Zielona G\'ora, Lubuska 2, 65-265 Zielona G\'ora, Poland}

\begin{document}

\date{Accepted 2011 November 24. Received 2011 November 23; in original form 2011 September 16}

\pagerange{\pageref{firstpage}-\pageref{lastpage}} \pubyear{YYYY}

\maketitle

\label{firstpage}

\begin{abstract}
We present fast timing photometric observations of the intermediate polar V2069 Cygni (RX J2123.7+4217) using the Optical Timing Analyzer (OPTIMA) at the 1.3 m telescope of Skinakas Observatory. The optical (450-950 nm) light curve of V2069 Cygni was measured with sub-second resolution for the first time during July 2009 and revealed a double-peaked pulsation with a period of $743.38\pm 0.25$. A similar double-peaked modulation was found in the simultaneous \textit{Swift} satellite  observations. We suggest that this period represents the spin of the white dwarf accretor. Moreover, we present the results from a detailed analysis of the \textit{XMM--Newton} observation that also shows a double-peaked modulation, however shifted in phase, with $742.35\pm 0.23$~s period. The X-ray spectra obtained from the \textit{XMM--Newton} EPIC (European Photon Imaging Camera) instruments were modelled by a plasma emission and a soft black body component with a partial covering photo-electric absorption model with covering fraction of 0.65. An additional Gaussian emission line at 6.385~keV with an equivalent width of 243~eV is required to account for fluorescent emission from neutral iron. The iron fluorescence ($\sim$6.4~keV) and FeXXVI lines ($\sim$6.95~keV) are clearly resolved in the EPIC spectra. In the \textit{P$_{\rm orb}$--P$_{\rm spin}$} diagram of IPs, V2069~Cyg shows a low spin to orbit ratio of $\sim$0.0276 in comparison with $\sim$0.1 for other intermediate polars. 
\end{abstract}


\section{Introduction}
Magnetic cataclysmic variables (CVs) are interacting close binary systems in which material transferred from a Roche-lobe filling low mass companion is accreted by a magnetic white dwarf (WD). Magnetic CVs are subdivided in two groups: polars (or AM Her type) and intermediate polars (IPs; or DQ Herculis type). In polars, the WD has a sufficiently strong magnetic field \textit{(B~$\sim 10^{7}$$-$$10^{8}$}~G) which locks the system into synchronous rotation \textit{(P$_{\rm{spin}}$ = P$_{\rm{orb}}$)} and prevents the accretion disk to form around the WD. In IPs, the field of the WD is one order of magnitude weaker \textit{(B~$\sim 10^{6}$$-$$10^{7}$}~G), and therefore insufficient to force the WD to spin with the same period as the binary system orbits \textit{(P$_{\rm{spin}}$ $<$ P$_{\rm{orb}}$)}. The accretion in IPs happens through a disk with a disrupted inner region \citep{Cropper1990,Patterson1994,Warner1995,Hellier2001}.
\\

V2069~Cyg (RX J2123.7+4217) was discovered as a hard X-ray source by \citet{Motch-etal1996} and identified as a CV. \citet{ThorstensenTaylor2001} reported a most probable orbital period of 0.311683 days (7.48 h) from their spectroscopic observations. \cite{deMartino-etal2009} performed a preliminary analysis of \textit{XMM--Newton} observations that showed a strong peak at the fundamental frequency of 116.3 cycles d$^{-1}$ and  harmonics up to the third in the power spectrum. Additionally, the sinusoidal fit to the profile from both EPIC-pn and EPIC-MOS data revealed a fundamental period of $743.2\pm 0.4$ s and 55 per cent pulsed fraction. They also reported a spectral fit consisting of a 56~eV black body (bbody) component plus 16~keV thermal plasma emission and a Gaussian at 6.4~keV emission line with an equivalent width (EW) of 159~eV, being absorbed by a partial (69 per cent) covering model with $N_{\rm{H}} = 1.1\times 10^{23} ~\rm{cm^{-2}}$ and a total absorber with $N_{\rm{H}} = 5\times 10^{21} ~\rm{cm^{-2}}$. Their spectral analysis confirmed that V2069~Cyg is a hard X-ray emitting IP with a soft X-ray component. \citet{Butters-etal2011} carried out an analysis of \textit{RXTE} data   in the $2.0$$-$$10.0~\rm{keV}$ energy range and found the spin period of the V2069~Cyg WD to be $743.2 \pm 0.9~\rm{s}$ with a double-peak modulation. They also reported the spectral results with a 6.4~keV iron line which is typical of IPs.

\section{Observations} 
\subsection{High time resolved photometric observations}
We performed photometric observations of V2069~Cyg with the Optical Timing Analyzer (OPTIMA) instrument at the 1.3~m telescope at Skinakas Observatory, Crete, Greece. The high-speed photometer OPTIMA is a sensitive, portable detector to observe extremely faint optical pulsars and other highly variable astrophysical sources. The detector contains eight fibre-fed single photon counters-avalanche photo-diodes (APDs), and a GPS (Global Positioning System) for the time control. Single photons are recorded in all channels with absolute time tagging accuracy of $\sim 4~\rm{\mu s}$. The quantum efficiency of the APDs reaches a maximum of 60$\%$ at 750 nm and lies above 20$\%$ in the range 450--950~nm \citep{Kanbach-etal2003}. To observe V2069~Cyg OPTIMA was pointed at RA(J2000) = $21^{\rm h} 23^{\rm m} 44 \fs 82$, Dec(J2000) = $+42\degr 18\arcmin 01 \farcs 7$, corresponding to the central aperture of  a hexagonal bundle of fibres (Fig.~\ref{Fig:1}). A separate fibre is located at a distance of $\sim 1 \arcmin$ as a night sky background monitor. The log of the observations is given in Tab.~\ref{Tab:1}. 

\subsection{\textit{Swift}/XRT observations}
The simultaneous soft X-ray observations of V2069~Cyg were performed with the \textit{Swift}'s X-ray telescope (XRT; Burrows etal. 2005) in the energy range of 0.3$-$10~\rm{keV}. The CCD of the \textit{Swift}/XRT was operated in the Photon-Counting mode which retains full imaging and spectroscopic resolution with a time resolution of 2.54~s. The \textit{Swift} source position is: RA(J2000) = $21^{\rm h} 23^{\rm m} 44 \fs 69$ Dec.(J2000) = $+42\degr 17\arcmin 59 \farcs 6$ with an error radius of $3 \farcs 5$. For the XRT data we applied the following types of filters: grade 0--4, and a circular region filter centred at the position of the source with 10-pixels radius (corresponding to $\sim 23 \farcs 5$).

\subsection{\textit{XMM--Newton} observations}
The \textit{XMM--Newton} observation of V2069~Cyg was performed on 2009 April 30 (Observation ID: 0601270101). The EPIC instruments were operated in full-frame imaging mode with thin and medium optical blocking filters for EPIC-pn \citep{Strueder-etal2001} and EPIC-MOS \citep{Turner-etal2001}, respectively. The exposure times were 26433~s for EPIC-pn, 28023~s for EPIC-MOS1, 28029~s for EPIC-MOS2. We used the \textit{XMM--Newton} Science Analysis Software (SAS) v.10.0.0 to process the event files. The source coordinates derived from a standard source detection analysis of the combined EPIC images are RA(J2000) = $21^{\rm h} 23^{\rm m} 44 \fs 60$, Dec(J2000) = $+42\degr 18\arcmin 00 \farcs 1$. We identified the circular photon extraction regions (with radius of $36\arcsec$, $53\arcsec$ and $56\arcsec$ for EPIC-pn, EPIC-MOS1 and EPIC-MOS2 respectively) around the source by optimising the signal to noise ratio. A circular region was used for the background extraction from a nearby source-free area (with radius of $35\arcsec$) on the same CCD as the source. To create spectra we selected single-pixel events (PATTERN=0) from EPIC-pn data and single- to quadruple-pixel events (PATTERN 0--12) from EPIC-MOS data. For the timing analysis we used single- and double-pixel events from the EPIC-pn data (PATTERN 0--4), and single- to quadruple-pixel events from EPIC-MOS data. We sorted out bad CCD pixels and columns (FLAG=0). After the standard pipeline processing of the EPIC photon event files, we rejected some part of the data which was affected by very high soft proton background. We created good time intervals (GTIs) from background light curves ($7.0$$-$$15.0~\rm{keV}$ band) using count rates below 15 cts ks$^{-1}$ arcmin$^{-2}$ for EPIC-pn data and 2.5 cts ks$^{-1}$ arcmin$^{-2}$ for MOS data. The spectra of EPIC-pn, EPIC-MOS1 and EPIC-MOS2 contain 10576, 5908, and 6000 background subtracted counts, respectively.

\begin{table}
\caption{Log of the photometric ({OPTIMA}) and X-ray (\textit{Swift}/XRT and \textit{XMM--Newton}/EPIC) observations of V2069~Cyg.}
\label{Tab:1}
\begin{center}
\begin{tabular}{cccccc}
\hline
No. & Date & Detector & ObsBeg & Expo. \\
& 2009  & & (MJD) & (h) \\
\hline
1 & Jul 02 & OPTIMA & 55014.922 & 2.5 \\
2 & Jul 18 & OPTIMA & 55030.951 & 1.2 \\
3 & Jul 19 & OPTIMA & 55031.845 & 2.1 \\
4 & Jul 21 & OPTIMA & 55033.820 & 4.1 \\
5 & Jul 22 & OPTIMA & 55034.871 & 3.0 \\
6 & Jul 24 & OPTIMA & 55036.804 & 1.2 \\
7 & Jul 26 & OPTIMA & 55038.040 & 1.4 \\
8 & Jul 26 & OPTIMA & 55038.827 & 1.7 \\
9 & Jul 28 & OPTIMA & 55040.897 & 1.5 \\
A & Jul 21 & \textit{Swift} & 55033.786 & 0.8\\
B & Jul 22 & \textit{Swift} & 55034.048 & 0.9\\
C & Apr 30 & \textit{XMM--Newton} & 54951.463 & 7.8\\
\hline
\end{tabular}
\end{center}
\end{table}

\begin{figure}
\begin{center}
\includegraphics[width=6.5cm]{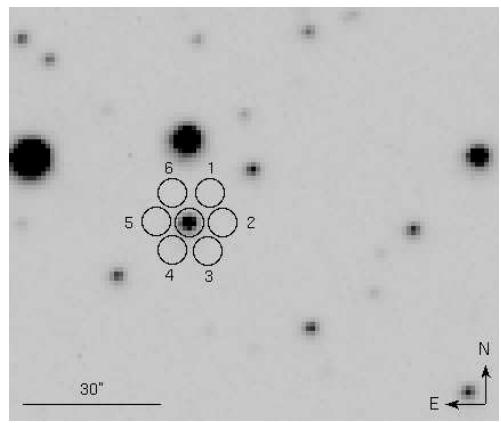}
\caption{OPTIMA fibre bundle centred on V2069~Cyg. The ring fibres (1--6) are used to monitor the background sky simultaneously.}
\label{Fig:1}
\end{center}
\end{figure}

\section[]{Data Analysis} 
\subsection[]{Timing analysis of the OPTIMA and \textit{Swift}/XRT data}
\begin{figure}
\centering
\includegraphics[width=7cm]{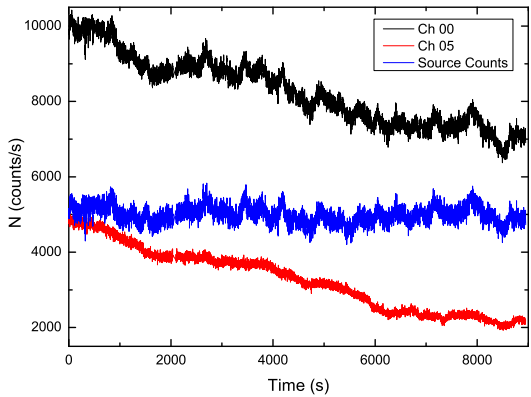}
\caption{OPTIMA light curves of V2069~Cyg from July 2nd, 2009 observation (no. 1 in Tab.~\ref{Tab:1}), shown as raw, uncalibrated countrates binned in 1~s intervals. Source count rates were obtained from the central fibre (channel 0) after subtraction of the properly calibrated sky background trace (channel 5). The sky background is decreasing in brightness because of the setting Moon.} 
\label{Fig:2}
\end{figure}

We analysed the data using the \texttt{HEASOFT} analysis package v.6.9. The X-ray and optical photon arrival times were converted to the solar system barycentre. OPTIMA count rates of the source were obtained from the central fibre (see Fig.~\ref{Fig:1}). Raw data were binned with 1~s and, after 'flat-fielding' all fibre channels on a source free region of sky background, the corresponding calibrated background counts were subtracted. We chose the fibre number 5 as the best representative of the background, because its APD response was closest to the APD response of channel 0 (Fig.~\ref{Fig:2}).

\begin{figure}
\begin{center}
\includegraphics[width=6.5cm, angle=270]{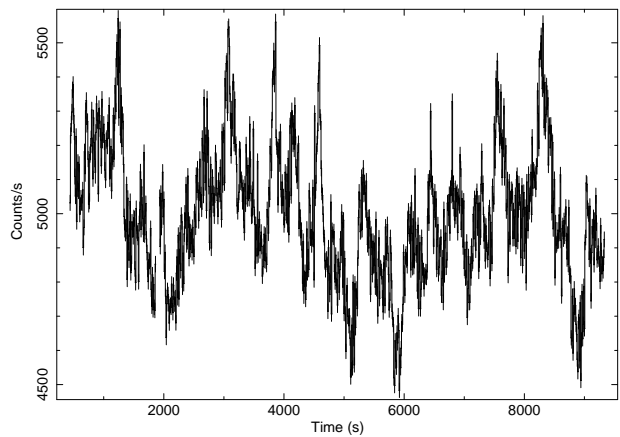}
\caption{Light curve of V2069~Cyg as derived in Fig.~\ref{Fig:2}, zoomed in the count rate scale for better visibility. The optical periodicity is clearly visible. The data are background subtracted and binned into 10~s intervals. Time 0 corresponds to MJD = 55014.92172.}
\label{Fig:3}
\end{center}
\end{figure}

\begin{figure}
\begin{center}
\includegraphics[width=6.5cm, angle=270]{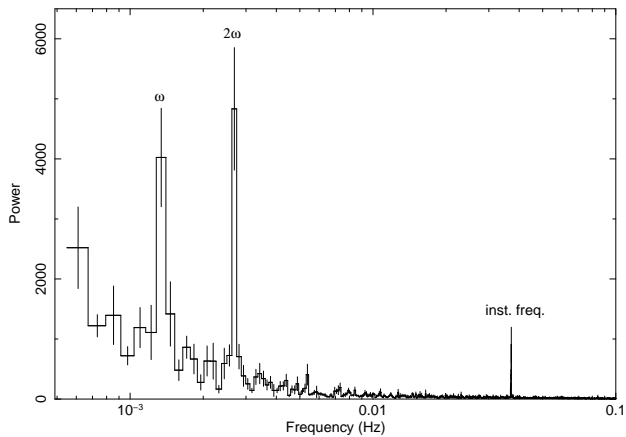}
\caption{Power spectrum obtained from OPTIMA data (Tab.~\ref{Tab:1}, all epochs). It shows prominent peaks at the fundamental spin frequency (first harmonic) of 0.00134277~Hz and its second harmonic of 0.00268555~Hz. An instrumental frequency at 0.0371094~Hz is also visible.}
\label{Fig:4}
\end{center}
\end{figure}

\begin{figure}
\begin{center}
\includegraphics[width=6.5cm, angle=270]{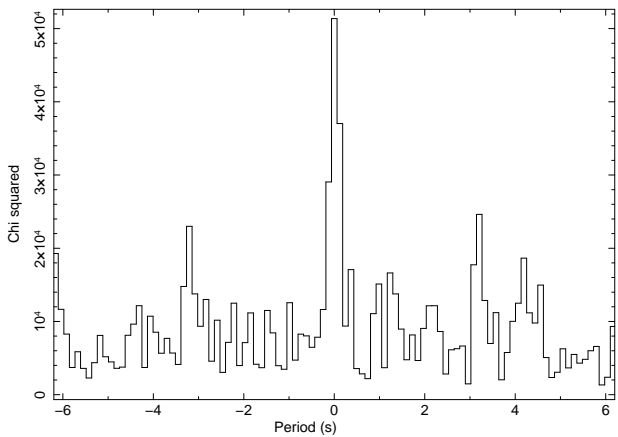}
\caption{$\chi^{2}$ periodogram as a function of the period, obtained from OPTIMA data. The central value (=0) corresponds to the best spin period of 743.38~s.}
\label{Fig:5}
\end{center}
\end{figure}

The resulting photometric light curve shows a prominent periodic variability (Fig.~\ref{Fig:3}). The power spectrum was computed with the Fast Fourier Transform (FFT) algorithm and normalised such that the white noise level expected from the data uncertainties corresponds to a power of 2 (Fig.~\ref{Fig:4}). The power spectrum shows peaks at the fundamental spin frequency (first harmonic) 0.00134277~Hz and its second harmonic 0.00268555~Hz (periods 744.73~s and 372.35~s, respectively), as well as a known systematic frequency of 0.03718~Hz (26.9~s). A $\chi^{2}$ folding analysis which folds the data over a range of periods reveals the best spin period of the WD as $743.38 \pm 0.25~\rm s$, Fig.~\ref{Fig:5}.

\begin{figure}
\begin{center}
\includegraphics[width=6.25cm, angle=270]{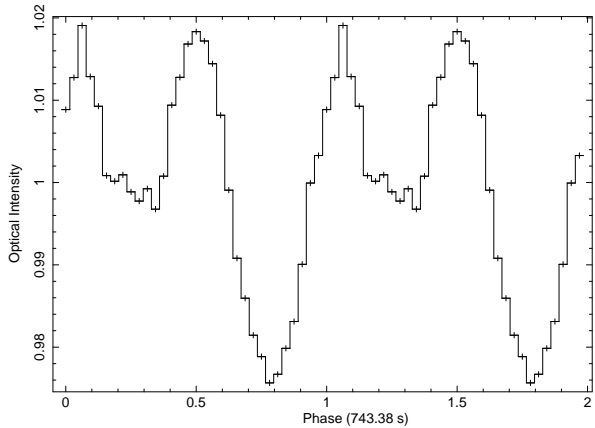}
\caption{Pulse profile obtained from all OPTIMA data (Tab.~\ref{Tab:1}, all epochs) folded with the 743.38~s spin cycle (32 bins/period). The profile is background subtracted and normalised to the average count rate of 4621 cts s$^{-1}$. Epoch, MJD = 54951.0.}
\label{Fig:6}
\end{center}
\end{figure}

\begin{figure}
\begin{center}
\includegraphics[width=6.5cm, angle=270]{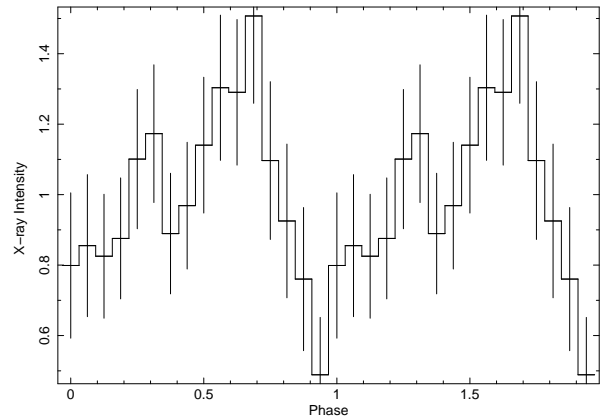}
\caption{Pulse profile obtained from \textit{Swift}-XRT data ($0.3$$-$$10.0~\rm{keV}$) folded at 743.38~s (16 bins/period) with an arbitrary zero point (MJD = 55014.0). The profile is background subtracted and normalised to an average count rate of 0.0918 cts s$^{-1}$.}
\label{Fig:7}
\end{center}
\end{figure}

The optical light curve folded with the 743.38~s spin period shows a double-peaked profile (see Fig.~\ref{Fig:6}) with very high duty cycle ($\sim 90$ per cent), that is the percentage of the rotation phase where there is a pulsed emission. Since the power spectrum of the optical data is dominated by the second harmonic of the spin frequency it is clearly seen that these two peaks are similar and separated by about half of the cycle in phase. \citet{Norton-etal1999} reported the same result in the X-ray data of IP YY~Dra, where the power spectrum is dominated by the second harmonic (i.e. $2/{\rm{P_{spin}}}$). On the other hand, due to low statistics, we could not determine the spin period from the \textit{Swift}-XRT data, therefore the XRT data were folded according to the optical period. The \textit{Swift}/XRT also shows a double-peak modulation at the WD spin period of 743.38~s (see Fig.~\ref{Fig:7}). However, the weaker peak is only marginally visible and is separated by less than half the pulse cycle.

\subsection[]{Timing analysis of the \textit{XMM--Newton} data}
\begin{figure}
\begin{center}
\includegraphics[width=6.5cm, angle=270]{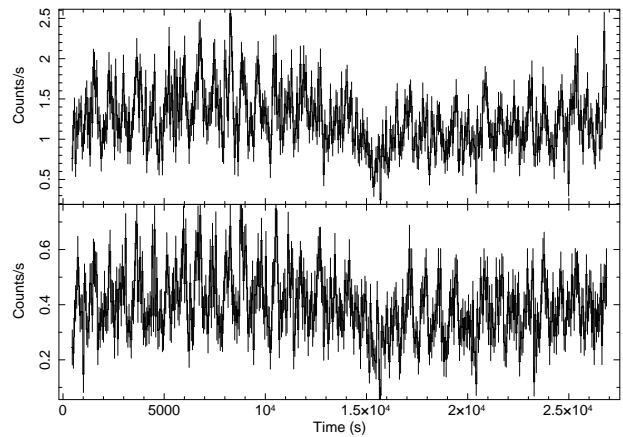}
\caption{X-ray broad-band (0.2$-$10.0~\rm{keV}) light curves of V2069~Cyg obtained from the EPIC-pn (top) and summed MOS1 and MOS2 (bottom) data. The periodic variations can be seen clearly. The data are background subtracted and binned to 55~s. Time 0 corresponds to MJD = 54951.46333.} 
\label{Fig:8}
\end{center}
\end{figure}

\begin{figure}
\begin{center}
\includegraphics[width=6.5cm, angle=270]{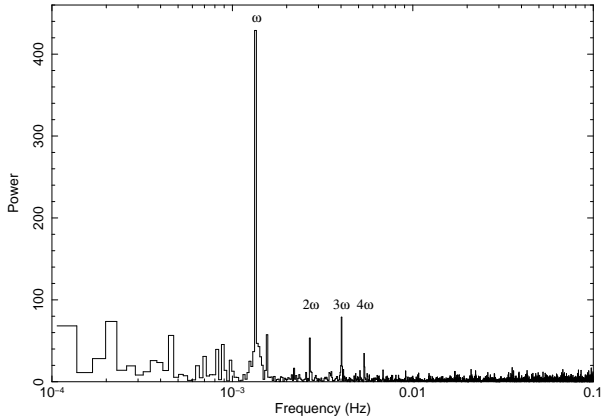}
\caption{Power spectrum obtained from the combined and merged event data of EPIC-pn and EPIC-MOS (0.2$-$10.0~\rm{keV}). It shows a strong peak at the fundamental frequency of 0.00134277~Hz which corresponds to the spin period of 744.73~s, and peaks at the second (0.00268555~Hz), third (0.00402832~Hz) and fourth (0.00537109~Hz) harmonic. The time binning of the input light curve is 1~s.}
\label{Fig:9}
\end{center}
\end{figure}

For the timing analysis of the \textit{XMM--Newton} data we corrected the event arrival times to the solar system barycentre. The background subtracted X-ray light curves in the 0.2$-$10.0~\rm{keV} energy band obtained from EPIC-pn and combined MOS data with a time binning of 55~s are shown in Fig.~\ref{Fig:8}. The periodic variations around 745 s can be seen clearly in the X-ray light curves. To improve the statistics for timing analysis a combined EPIC-pn, EPIC-MOS1 and EPIC-MOS2 event list from the source extraction region was created. The FFT timing analysis of the combined X-ray data revealed the presence of four harmonic frequencies with a strong peak at the fundamental frequency of 0.00134277~Hz that corresponds to a period of 744.73~s, as shown in Fig.~\ref{Fig:9}. We found that the fundamental frequency is much stronger than the second harmonic at energies above 0.5~keV, while the second harmonic (with very weak power) is stronger than the fundamental frequency at energies below 0.5~keV. A similar behaviour was also reported by \citet{EvansHellier2004} for V405~Aur. To determine the pulse period and its error we applied the Bayesian formalism as described in \citet{GregoryLoredo1996}. Using the combined and merged EPIC data in the 0.2$-$10~\rm{keV} energy band reveals the  spin period of the WD as $742.35\pm 0.23~\rm{s}$, 1$\sigma$ uncertainty. We obtained the optical spin period a bit longer than the X-ray spin period, however both periods are compatible within their errors.

\begin{figure}
\begin{center}
\includegraphics[width=6.5cm, angle=270]{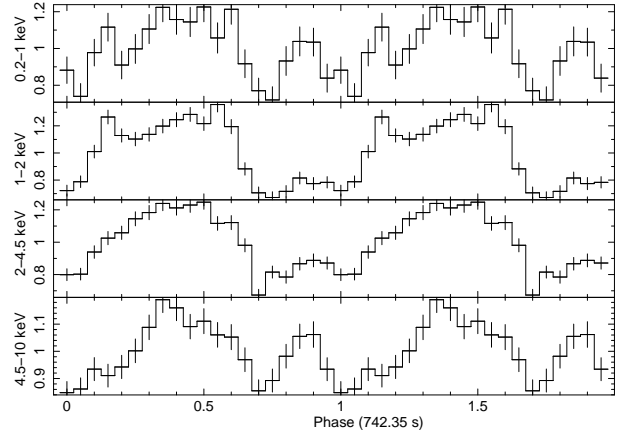}
\caption{Pulse profiles obtained from EPIC binned data folded with 742.35~s spin cycle (20 bins/period) for different energy ranges: 0.2$-$1.0~\rm{keV}, 1.0$-$2.0~\rm{keV}, 2.0$-$4.5~\rm{keV} and 4.5$-$10.0~\rm{keV} from top to bottom. The intensity profiles are background subtracted and normalised to average count rates of 0.078, 0.167, 0.221, and 0.196 cts~s$^{-1}$ (from top to bottom). Epoch, MJD = 54951.0.}
\label{Fig:10}
\end{center}
\end{figure}

\begin{figure}
\begin{center}
\includegraphics[width=6.5cm, angle=270]{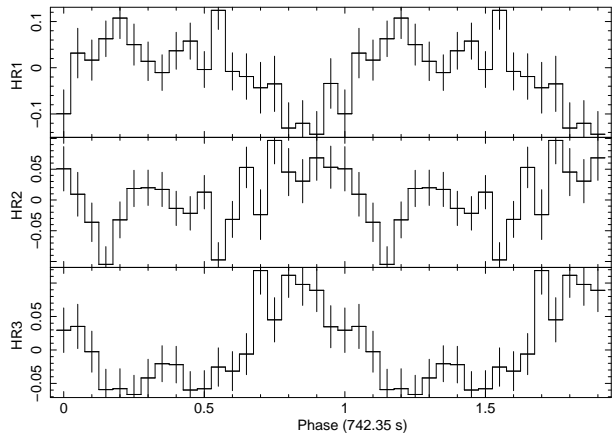}
\caption{Hardness ratio as a function of phase derived from the pulse profiles (Fig.~\ref{Fig:10}) in two neighbouring standard energy bands (0.2$-$1.0~\rm{keV} and 1.0$-$2.0~\rm{keV}, 1.0$-$2.0~\rm{keV} and 2.0$-$4.5~\rm{keV} and 2.0$-$4.5~\rm{keV} and 4.5$-$10.0~\rm{keV}, from top to bottom).}
\label{Fig:11}
\end{center}
\end{figure}

\begin{figure}
\begin{center}
\includegraphics[width=6.5cm, angle=270]{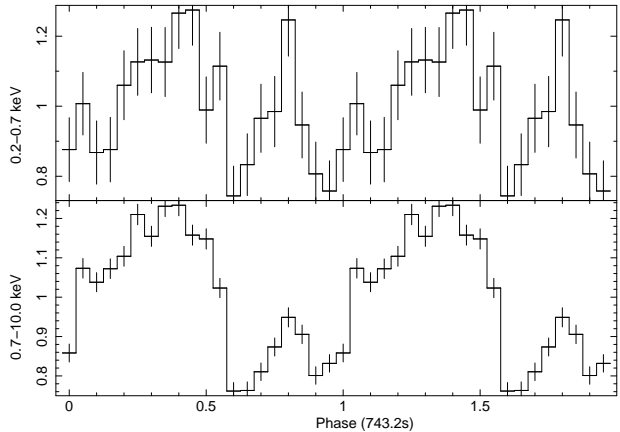}
\caption{Pulse profiles folded with 742.35~s (20 bins/period) obtained from combined EPIC data (pn, MOS1 and MOS2) in the energy range 0.2$-$0.7~\rm{keV} and 0.7$-$10.0~\rm{keV}. Epoch, MJD = 54951.0.}
\label{Fig:12}
\end{center}
\end{figure}

We folded the light curve to obtain the pulse profiles from the EPIC data (Fig.~\ref{Fig:10}) with the spin period in the different energy bands of 0.2$-$1.0~\rm{keV}, 1.0$-$2.0~\rm{keV}, 2.0$-$4.5~\rm{keV} and 4.5$-$10.0~\rm{keV} and calculated hardness ratios (Fig.~\ref{Fig:11}) as a function of pulse phase. The hardness ratios were derived from the pulse profiles in two neighbouring standard energy bands [HR$_{\rm{i}}$=(R\textit{$_{\rm{i+1}}-$R$_{\rm{i}}$})/(\textit{R$_{\rm{i+1}}$+R$_{\rm{i}}$}), where \textit{R$_{\rm{i}}$} denotes the background subtracted count rate in the energy band $i$, with $i$ from 1 to 4]. The \textit{XMM--Newton} data also show a double-peaked modulation with 742.35~s period consistent with the values obtained from OPTIMA, \textit{Swift}/XRT and \textit{RXTE} data. The double-peaked pulse profile is more prominent at lower energies (0.2$-$0.7~\rm{keV}), while the second peak is weaker at the higher energies (0.7$-$10.0~\rm{keV}; see Fig.~\ref{Fig:12}). Here the second peak is separated by less than half of the pulse cycle, and the power spectrum of the X-ray data is dominated by the fundamental spin frequency (i.e. \textit{1/P$_{\rm{spin}}$}). A similar behaviour was observed in the X-ray data of IP V709~Cas by \citet{Norton-etal1999}, where the power spectrum is dominated by the fundamental harmonic. The pulse profiles have highly asymmetrical rise and decay flanks. A dip feature is significant before the primary pulse maximum in the 0.2$-$1.0~\rm{keV} band and centred on the primary maximum in the 1.0$-$2.0~\rm{keV} band, while the primary maximum is more symmetric at higher energies (Fig.~\ref{Fig:10}). A similar feature was also observed in V709~Cas \citep{deMartino-etal2001}, in NY~Lup \citep{Haberl-etal2002} and in UU~Col \citep{deMartino-etal2006b}. The evolution of the pulse profiles with double-peaked structure from lower energies to higher, is causing the variations in the hardness ratios. In Fig.~\ref{Fig:11}, the hardness ratios show a hardening (increase) at spin minimum and a softening (decrease) at spin maximum which is more prominent in HR3. This typical behaviour is often observed from IPs and is generally produced by the larger photoelectric absorption when viewing along the accretion curtain \citep{deMartino-etal2001,Haberl-etal2002}. In HR2, the ratio shows two asymmetric maxima, separated by a dip centred on the primary spin maximum seen in the 1.0$-$2.0~\rm{keV} band and a second one appearing with a toothed-shape produced by the secondary spin maximum (see Fig.~\ref{Fig:10}). In  HR1, an antiphase behaviour is observed with respect to HR2.

\subsection[]{Orbital phase resolved timing analysis}
\begin{figure}
\begin{center}
\includegraphics[width=6.5cm, angle=270]{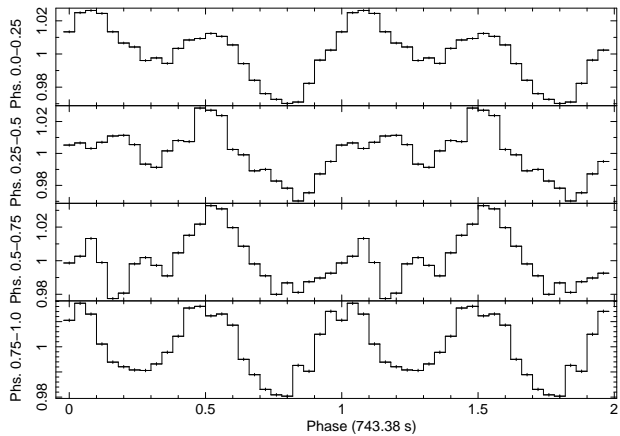}
\caption{Orbital phase resolved (0.0$-$0.25, 0.25$-$0.5, 0.5$-$0.75 and 0.75$-$1.0) pulse profiles folded with 743.38~s 
(25 bins/period) obtained from OPTIMA data (Tab.~\ref{Tab:1}, all epoch). Epoch, MJD = 54951.0.}
\label{Fig:13}
\end{center}
\end{figure}

\begin{figure}
\begin{center}
\includegraphics[width=6.5cm, angle=270]{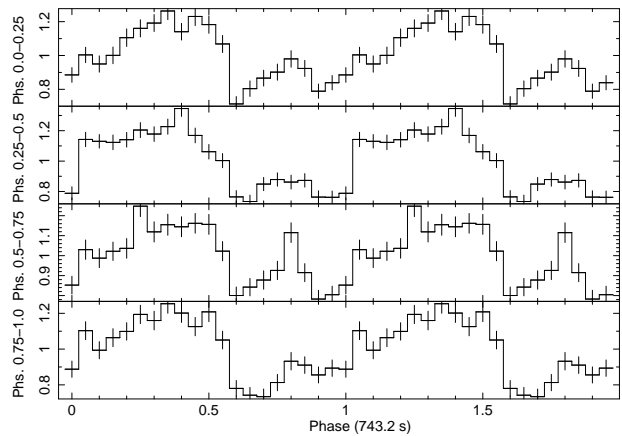}
\caption{Orbital phase resolved (0.0$-$0.25, 0.25$-$0.5, 0.5$-$0.75 and 0.75$-$1.0) pulse profiles folded with 742.35~s 
(16 bins/period) obtained from combined EPIC data (pn, MOS1 and MOS2). Epoch, MJD = 54951.0.}
\label{Fig:14}
\end{center}
\end{figure}

We investigated if the pulse shape of the rotating WD is changing with orbital phase of the binary system. The orbital phase was determined with the following ephemeris: phase (BJD)= $[T-2451066.7837(20)]/0.311683(2)$, where, T is the observation time \citep{ThorstensenTaylor2001}. For this purpose we obtained the WD pulse profiles in four orbital phase ranges: 0.0$-$0.25, 0.25$-$0.5, 0.5$-$0.75 and 0.75$-$1.0. Results are shown in the Fig.~\ref{Fig:13} and Fig.~\ref{Fig:14} for OPTIMA and EPIC data, respectively. There is some indication of a profile change, especially in the orbital phase range 0.5$-$0.75, for both optical and X-ray light curves.

\subsection[]{Spectral analysis of the \textit{XMM--Newton} data}
\begin{figure}
\begin{center}
\includegraphics[width=6.5cm, angle=270]{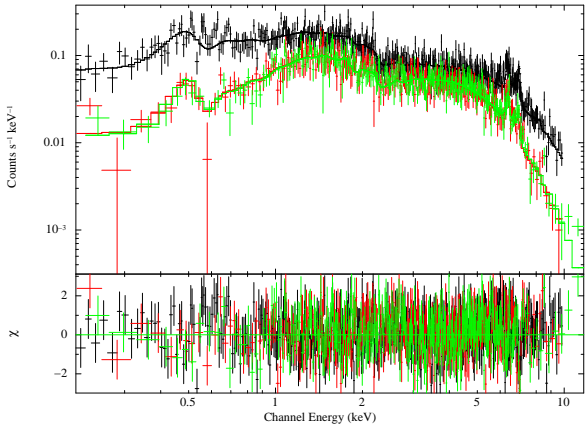}
\caption{The composite model (phabs*pcfabs* (mekal + bbody + gaussian)*constant) fitted to the spectrum of the EPIC-pn (black) and MOS (green and red) data in the 0.2$-$10~\rm{keV} energy band. The bottom panel shows the residuals.}
\label{Fig:15}
\end{center}
\end{figure}

In order to estimate the basic parameters of the emitting region, a spectral analysis of the X-ray data was performed with \texttt{XSPEC} v.12.5.0x \citep{Arnaud1996}. The three EPIC spectra were fitted simultaneously with a model consisting of thermal plasma emission (\texttt{MEKAL}; \citet{Mewe-etal1985}) and a soft bbody component \citep[as suggested by ][]{deMartino-etal2009}, absorbed by a simple photoelectric absorber (phabs) and a partially-covering photoelectric absorber (pcfabs). An additional Gaussian line is required which represents iron K fluorescent emission at 6.4~keV as is often seen from classical IPs. To account for cross-calibration uncertainties a constant factor was introduced. The absorbers phabs and pcfabs describe the absorptions of the interstellar (along the line of sight) and circumstellar (inside the system by the accretion curtain/stream), respectively \citep{Staude-etal2008}. The MEKAL model produces an emission spectrum from hot diffuse gas above the WD's surface and includes line emissions from various elements. In a first fit to the spectra, the plasma temperature for the MEKAL component could not be constrained. Therefore, we fixed the plasma temperature at 20~keV, a value typical for IPs \citep{Staude-etal2008}. We obtained a best fit with reduced $\chi^{2}_r$ of 1.002 ($\chi^{2}$ of 1014.69 with 1013 degrees of freedom). The spectral parameters for the fit are summarised in Tab.~\ref{Tab:2} and the spectra including the best fit model is shown in Fig.~\ref{Fig:15}. 

\begin{figure}
\begin{center}
\includegraphics[width=6.5cm, angle=270]{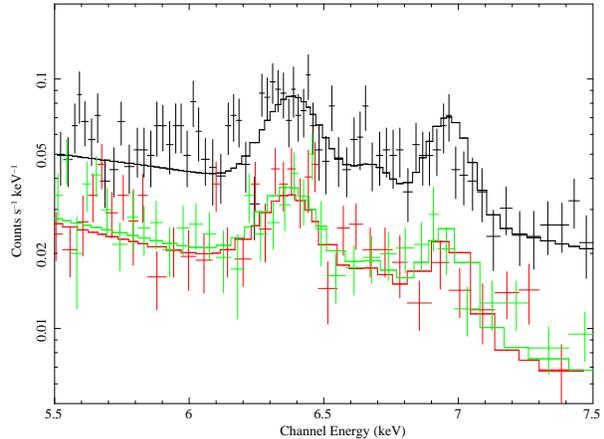}
\caption{Enlarged part of Fig.~\ref{Fig:15} showing the Fe line complex in the EPIC spectra.}
\label{Fig:16}
\end{center}
\end{figure}

We determined the hydrogen column density as $N_{\rm H} = 3.84\times 10^{21}~\rm{cm^{-2}}$. This is higher than the total Galactic hydrogen column density ($3.79\times 10^{20}~\rm{cm^{-2}}$, an interpolated value from \citet{Dickey-etal1990} that was calculated using the HEASARC $N_{H}$ web interface\footnote{http://heasarc.nasa.gov/cgi-bin/Tools/w3nh/w3nh.pl}) in the direction of the source. Our result is comparable to the value ($5\times 10^{21}~\rm{cm^{-2}}$) obtained by \citet{deMartino-etal2009}. For the partial absorber we find, $N_{\rm H} = 8.29\times 10^{22}~\rm~cm^{-2}$ with a covering fraction of 0.65. Similar values of partial absorber were derived for V2069~Cyg \citep{deMartino-etal2009} and the other soft IPs observed with \textit{XMM--Newton} (see Tab.~\ref{Tab:2a}). The absorbed flux of V2069~Cyg in the 0.2$-$10.0~\rm{keV} energy band (derived for EPIC-pn) is $7.93 \times 10^{-12}~\rm ergs~cm^{-2}s^{-1}$ which corresponds to a source intrinsic flux (with absorption set to 0) of $2.64\times 10^{-11}~\rm{ergs~cm^{-2}s^{-1}}$ (EPIC-MOS values are 2 per cent higher corresponding to the constant factors derived from the fit). The spectra around the Fe-K emission line complex are shown enlarged in Fig.~\ref{Fig:16}. The iron fluorescence and Fe$\texttt{XXVI}$ lines are clearly resolved in the EPIC spectra. The Fe$\texttt{XXVI}$ line energy identified from the \texttt{XSPEC} possible lines list is $\sim$6.95~keV and the fluorescence line energy derived from the fit is $\sim$6.385 $\pm$0.017 keV. The EW of the fluorescent line is 243~eV.

\begin{table*}
\begin{center}
\begin{minipage}{115mm}
\caption{Spectral fit result for the \textit{XMM--Newton} EPIC data.}
\label{Tab:2}
\begin{tabular}{@{}llccc@{}}
\hline
Model & Parameter & Unit  & Value & error \\
\hline                        
phabs   &  \textit{NH} & 10$^{21}$ cm $^{-2}$  &  3.84 & (-0.04, +0.05)  \\
pcfabs  &  \textit{NH} & 10$^{22}$ cm $^{-2}$  &  8.29 & (-1, +1.2) \\
        &  CvrFract    &              &  0.65  & $\pm$0.02  \\
mekal   &  kT          & keV          &  20.0  & frozen 	 \\
        &  \textit{nH} & cm$^{-3}$    &  1.0   & frozen 	 \\
	&  Abundance   &              &  1.0   & frozen 	 \\
	&  norm        &              &  6.29$\times$10$^{-3}$ & (-2.4, +2.6)$\times$10$^{-4}$  \\ 
bbody   & kT           & keV          &  7.68$\times$10$^{-2}$ & (-4.3, +4.2)$\times$10$^{-3}$ \\ 
        & norm         &              &  2.18$\times$10$^{-4}$ & (-0.75, +1.2)$\times$10$^{-4}$ \\
gaussian &  LineE      & keV          &  6.385 & $\pm$0.017      \\
         &  Sigma      & eV           &  51 & (-32,  +27)        \\
	 &  norm       &              &  2.6$\times$10$^{-5}$ & (-5.4, +3.9) $\times$10$^{-6}$  \\
constant factor &  pn         &       &  1.0     & frozen \\
	 &  MOS1       &              &  1.026   & $\pm$0.018 	 \\
  	 &  MOS2       &              &  1.028   & $\pm$0.018 	 \\
\hline
\end{tabular}
\end{minipage}
\end{center}
\end{table*}

\begin{table}
\centering
\begin{minipage}{80mm}
\caption{The parameters of the partial absorber obtained for V2069 Cyg and some soft IPs observed with \textit{XMM--Newton}.}
\label{Tab:2a}
\begin{tabular}{@{}llrrrrlrlr@{}}
\hline
Source & $N_{\rm H}(cm^{-2})$ & CvrFract & Reference\footnote{References: (1) \citet{deMartino-etal2009}; (2) \citet{Staude-etal2008}; 
(3) \citet{Evans-etal2006}; (4) \citet{deMartino-etal2006b}; (5) \citet{EvansHellier2004}; (6) \citet{Haberl-etal2002};}\\
\hline
V2069~Cyg& $11\times 10^{22}$  & $0.69$ & 1 \\
MU~Cam   & $7.9\times 10^{22}$ & $0.61$ & 2 \\
PQ~Gem   & $11.1\times 10^{22}$& $0.45$ & 3 \\
UU~Col   & $10\times 10^{22}$  & $0.51$ & 4 \\
V405~Aur & $6.1\times 10^{22}$ & $0.52$ & 5 \\
NY~Lup   & $9.7\times 10^{22}$ & $0.47$ & 6 \\
\hline
\end{tabular}
\end{minipage}
\end{table}

\section[]{Discussion} 
We have presented the optical (OPTIMA) and X-ray (\textit{Swift}-XRT and \textit{XMM--Newton} EPIC) observations of the IP V2069~Cyg. The timing analysis of the optical and X-ray data reveals pulsations at periods of $743.38\pm 0.25~\rm{s}$ and $742.35\pm 0.23~\rm{s}$ respectively, representing the spin period of the WD. We have found that the second harmonic is much stronger than the fundamental in the power spectrum obtained from the optical data. Furthermore, the fundamental frequency from XMM data is weak or even absent at energies $<$ 0.5~keV, while it is stronger at $>$ 0.5~keV, compared to the second harmonic. IP V405~Aur has shown very similar behaviour in the \textit{XMM--Newton} data \citep{EvansHellier2004}. The double-peaked pulsations at the spin period are clearly observed in the optical and X-ray data (0.2$-$10~\rm{keV}). The folded light curves show a more prominent double-peaked pulse profile when the power spectrum is dominated by the second harmonic. When the second harmonic is weak the curve possesses a similar profile but with a weaker second peak. Therefore, the power spectrum of the optical data is dominated by the second harmonic, while the X-ray data is the fundamental. The peak separation is around 0.5 for the optical data, and less than 0.5 for the X-ray data. 

IP systems (assuming equilibrium rotation) with a short spin period will have relatively small magnetospheres, corresponding to the shorter Keplerian periods in the inner accretion disk. In such short period systems the WD is therefore expected to have a weak magnetic field. The magnetic forces of the WD pick up the material from the accretion disk approximately at the co-rotation radius. The material attaches to the field lines and is channelled onto the WD magnetic poles, where it undergoes a strong shock. After that it is settling on the surface and cooling by the emissions of X-ray bremsstrahlung and optical/infrared cyclotron \citep{Rosen-etal1988,Norton-etal2004a}. As proposed by \citet{EvansHellier2004} most likely the prominent double-peaked modulation in the soft X-ray emission is due to the changing viewing geometry onto the accreting polar caps. We view the heated surface of the WD most favourably when one of the poles points towards us. Nevertheless, due to the highly inclined dipole axis, the external regions of the accretion curtains will not quite cross the line of sight, therefore the hard X-ray emission exhibits a double-peaked pulse profiles with a weaker secondary peak. However, the intensity of the pulse profiles could be also affected by the opacity resulting in electron scattering and absorption in the highly ionized post-shock region, or an offset of the magnetic axis from the WD centre \citep{Allan-etal1996,Norton-etal1999,EvansHellier2004}. 

\begin{figure}
\begin{center}
\includegraphics[width=6.5cm, angle=270]{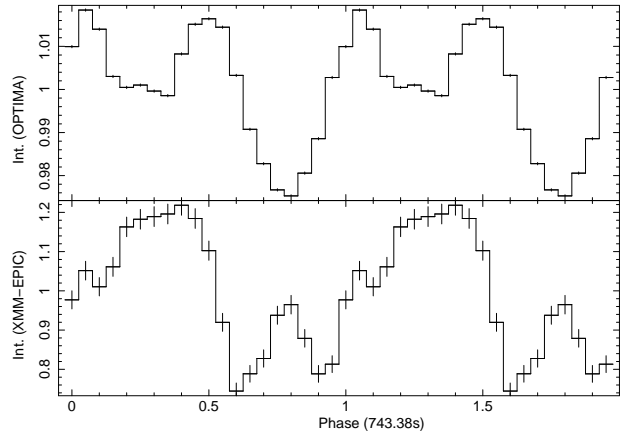}
\caption{Pulse profiles folded with 743.38~s (20 bins/period) obtained from combined XMM EPIC 0.2$-$10~\rm{keV} data and OPTIMA data. Epoch, MJD = 54951.0.}
\label{Fig:17}
\end{center}
\end{figure}

The pulse profiles of the optical and X-ray data (each folded with both 742.35~s and 743.38~s spin period and with a same reference time) are out of phase. As an example we show these profiles folded with 743.38~s in Fig.~\ref{Fig:17}. \citet{deMartino-etal2009} also reported that the X-ray pulses (from EPIC-pn) are anti-phased with the optical pulses (in the B-band from optical monitor on the XMM-Newton). X-ray and optical/infrared photons in some IPs originate from two different regions. The optical/infrared photons are thought to originate in the X-ray heated magnetic polar caps, and possibly in the accretion stream \citep{Eracleous-etal1994,Israel-etal2003,Revnivtsev-etal2010}. \citet{Norton-etal2004b} suggested that one of the magnetic poles heated by the accretion flow will leave behind a heated trail on the WD surface which will emit optical/infrared photons. During the emission some part of the optical/infrared photons will be absorbed by the flow while the accretion flow is heating the second pole. At that time the rest of the optical/infrared modulations will be seen which is shifted with respect to the X-rays. The phase shift observed between the optical and X-ray pulse profiles in V2069 Cyg is most probably caused by this X-ray heated mechanism.

On the other hand, \citet{Norton-etal1999} suggested that IPs which show a single-peaked pulse profile resulting from stream-fed (or disc-overflow) accretion are an indicator of a WD with a relatively strong magnetic field. These IPs with long WD spin periods (longer than 700 s) might show X-ray beat periods (1/\textit{P$_{\rm{beat}}$=}1/\textit{P$_{\rm{spin}}$}--1/\textit{P$_{\rm{orb}}$}) at some time in their lives (FO Aqr, TX Col, BG CMi, AO Psc, V1223 Sgr and RX J1712.6-2414). Conversely, IPs with short WD spin periods (shorter than 550~s) have shown double-peaked pulse profiles and must therefore have weak magnetic fields. These short period systems did not exhibit X-ray beat periods (AE Aqr, DQ Her, XY Ari, V709 Cas, GK Per, YY Dra and V405 Aur). In the power spectrum of V2069~Cyg we have not found any specific signal at the beat frequency. This absence indicates that in these short period IPs and V2069 Cyg accretion does not occur in a stream-fed or disc-overflow scenario \citep{Norton-etal1999}.

The X-ray spectra of V2069~Cyg can be described by thermal plasma emission (kT of $\sim 20$ keV) plus a soft bbody component with complex absorption and an additional fluorescent iron-K emission line, which originates on the WD surface (at 6.4~keV, with an EW of 243~eV). V2069~Cyg and V405~Aur show similar bbody parameters with kT of $\sim$77~eV and $\sim$40~eV \citep{EvansHellier2004}, respectively. Moreover, the two IPs have quite similar spin-orbit period ratios of 0.0276 for V2069~Cyg (743.38~s/26928~s) and 0.036 for V405~Aur (545.5~s/14986~s).

\begin{figure*}
\begin{center}
\includegraphics[width=12cm]{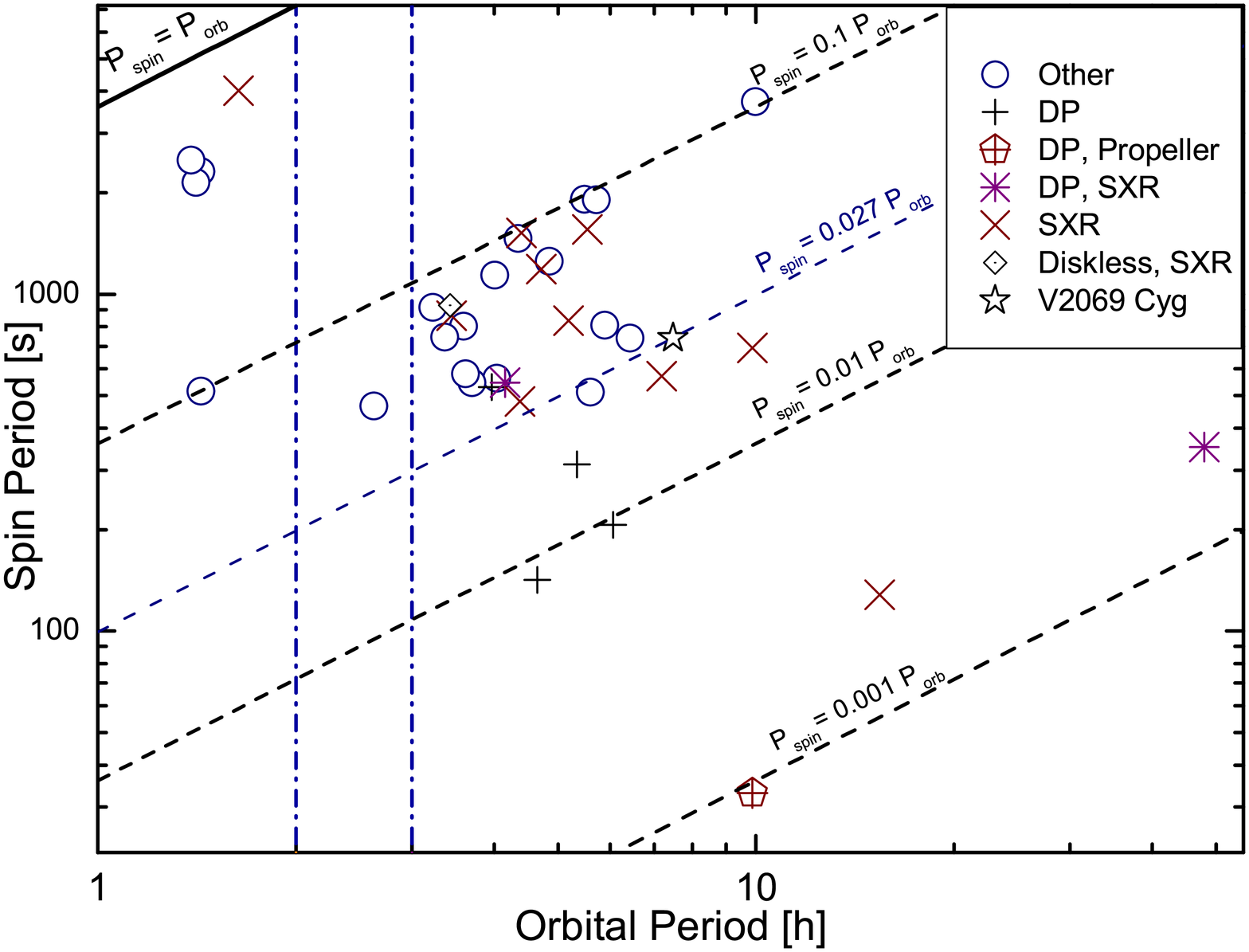}
\caption{\textit{P$_{\rm{orb}}$--P$_{\rm{spin}}$} diagram of 39 IPs: DP, double-peak pulsation; SXR, soft X-ray component; disk-less, have no accretion disk. The vertical dashed lines show the approximate location of the \textquoteleft period gap\textquoteright, and the diagonal lines are for \textit{P$_{\rm{spin}}$=P$_{\rm{orb}}$} (solid) and \textit{P$_{\rm{spin}}$ = 0.1/0.01/0.001 $\times~P_{\rm{orb}}$} (dashed). V2069~Cyg (shown with a star) is located well within the population of double-peak IPs with soft X-ray component but has a rather low spin-orbit period ratio of 0.0276.}
\label{Fig:18}
\end{center}
\end{figure*}

We adopted Mukai\textquoteright s classification\footnote{http://asd.gsfc.nasa.gov/Koji.Mukai/iphome/iphome.html} of IPs and updated his \textit{P$_{\rm{spin}}$--P$_{\rm{orb}}$} diagram to include V2069~Cyg (see  Fig.~\ref{Fig:18} and Tab.~\ref{Tab:3}). Several IPs are found close to \textit{P$_{\rm{spin}}/P_{\rm{orb}}$} = 0.1. There are 28 systems in the range of 0.01 \textit{$<$ P$_{\rm{spin}}/$P$_{\rm{orb}}$ $\leq$} 0.1 and \textit{P$_{\rm{orb}}$ $>$} 3 h, 5 systems with \textit{P$_{\rm{spin}}/$P$_{\rm{orb}}$ $\geq$} 0.1 and \textit{P$_{\rm{orb}}$ $<$} 2 h, and only one system with \textit{P$_{\rm{spin}}/$P$_{\rm{orb}}$ $\sim$} 0.049 that lies in the \textquoteleft period gap\textquoteright. Finally, there are 5 systems with \textit{P$_{\rm{spin}}$/P$_{\rm{orb}}$ $<$} 0.01. Those are defined as fast rotating WDs. Only one of them, AE Aqr, shows propeller behaviour. They also show the soft X-ray bbody component in their spectrum \citep{Norton-etal2004c,Parker-etal2005,EvansHellier2007,NortonMukai2007,Anzolin-etal2008}.

\begin{table*}
\centering
\begin{minipage}{160mm}
\caption{39 IPs with known spin and orbital periods.}
\label{Tab:3}
\begin{tabular}{@{}llrrrrlrlr@{}}
\hline
Name & P$_{\rmn{orb}}$ & P$_{\rmn{spin}}$ & P$_{\rmn{spin}}$ / P$_{\rmn{orb}}$ & Properties \footnote{SXR: soft X-ray bbody components; DP: double-peaked pulse profiles.} & Period References \footnote{REFERENCES: (1) \citep{Allan-etal1998}; (2)
\citep{Anzolin-etal2008};        (3) \citep{Baskil-etal2005};     (4) \citep{Kim-etal2005};
(5) \citep{Butters-etal2007};    (6) \citep{Bonnet-etal2007};     (7) \citep{Bonnet-etal2009};
(8) \citep{Buckley-etal1995};    (9) \citep{Burwitz-etal1996};    (10) \citep{Butters-etal2008};
(11) \citep{Butters-etal2009a};  (12) \citep{Butters-etal2009b};  (13) \citep{Choi-etal1999};
(14) \citep{Crampton-etal1986};  (15) \citep{deMartino-etal2008}; (16) \citep{deMartino-etal1999};
(17) \citep{deMartino-etal2006a};(18) \citep{deMartino-etal2009}; (19) \citep{Duck-etal1994};
(20) \citep{EvansHellier2004};   (21) \citep{EvansHellier2007};   (22) \citep{Gaensicke-etal2005};
(23) \citep{HarlaftisHorne1999}; (24) \citep{Haswell-etal1997};   (25) \citep{Hilton-etal2009}; 
(26) \citep{Hellier2007};        (27) \citep{Hellier-etal1994};   (28) \citep{Hellier-etal1997}; 
(29) \citep{Hellier-etal2002a};  (30) \citep{Hellier-etal2002b};  (31) \citep{Kemp-etal2002}; 
(32) \citep{Mauche2004};         (33) \citep{Norton-etal1999};    (34) \citep{Norton-etal2002};
(35) \citep{Patterson-etal2004}; (36) \citep{Pretorius2009};      (37) \citep{Hellier-etal1998}; 
(38) \citep{Staude-etal2003};    (39) \citep{Schlegel2005};       (40) \citep{Scaringi-etal2011}; 
(41) \citep{Scaringi-etal2010};  (42) \citep{Thorstensen-etal2010}; (43) \citep{deMartino-etal2006b}; 
(44) \citep{Patterson-etal2011}; (45) \citep{Zhang-etal1995};}\\  & (h) & (s) & & & \\
\hline
\textit{0.01 $<$ P$_{\rmn{spin}}$ / P$_{\rmn{orb}}$ $\leq$ 0.1} & \textit{and P$_{\rmn{orb}}$ $>$ 3 h} \\
\hline
V709 Cas & 5.341 & 312.780 & 0.01627 & DP & 33,41 \\
NY Lup & 9.870 & 693.010 & 0.01950 & SXR & 2, 21, 17, 41 \\
RXS J213344.1+510725 & 7.193 & 570.800 & 0.02204 & SXR & 2, 11, 42 \\
Swift J0732.5-1331 & 5.604 & 512.420 & 0.02540 & - & 5, 41 \\
V2069~Cyg & 7.480 & 743.384 & 0.02756 & SXR, DP & 18, 41 \\
RXS J070407.9-262501 & 4.380 & 480.670 & 0.03205 & SXR & 2, 22,44 \\
El Uma & 6.430 & 741.660 & 0.03204 & - & 3 \\
V405 Aur & 4.160 & 545.456 & 0.03642 & SXR, DP & 2, 20, 23, 41 \\
YY Dra & 3.969 & 529.310 & 0.03705 & DP & 24, 33 \\
IGR J15094-6649 & 5.890 & 809.700 & 0.03819 & - & 12, 41 \\
IGR J00234+6141 & 4.033 & 563.500 & 0.03881 & - & 6 \\
RXS J165443.5-191620 & 3.700 & 546.000 & 0.04099 & - & 40 \\
IGR J16500-3307 & 3.617 & 579.920 & 0.04454 & - & 36, 41 \\
PQ Gem &  5.190 & 833.400 & 0.04461 & SXR & 2, 19, 21, 27, 41 \\
V1223 Sgr & 3.366 & 745.630 & 0.06153 & - & 11, 41 \\
AO Psc & 3.591 & 805.200 & 0.06229 & - &  11, 41 \\
UU Col & 3.450 & 863.500 & 0.06952 & SXR & 2, 9, 21, 43 \\
MU Cam & 4.719 & 1187.250 & 0.06989 & SXR & 2,38, 41 \\
FO Aqr & 4.850 & 1254.400 & 0.07184 & - & 11, 16, 41 \\
V2400 Oph & 3.430 & 927.660 & 0.07513 & SXR, Diskless & 2, 8, 21, 26 \\
WX Pyx &  5.540 & 1557.300 & 0.07808 & SXR & 2, 21,39 \\
BG Cmi & 3.230 & 913.000 & 0.07852 & - & 4, 41 \\
IGR J17195-4100 & 4.005 & 1139.500 & 0.07902 & - & 36, 41 \\
TX Col & 5.718 & 1910.000 & 0.09284 & - & 41 \\
V2306 Cyg & 4.350 & 1466.600 & 0.09365 & - & 11, 34 \\
RXS J180340.0-401214 & 4.402 & 1520.510 & 0.09595 & SXR & 2, 22 \\
TV Col & 5.486 & 1911.000 & 0.09676 & - & 41 \\
V1062 Tau & 9.982 & 3726.000 & 0.10368 & - & 30, 42 \\
\hline
\textit{P$_{\rmn{spin}}$ / P$_{\rmn{orb}}$ $\geq$ 0.1} & \textit{and P$_{\rmn{orb}}$ $<$ 2 h} \\
\hline
HT Cam & 1.433 & 515.0592 & 0.09984 & - & 31 \\
V1025 Cen &  1.410 & 2147.000 & 0.42297 & - & 29, 37 \\
DW Cnc & 1.435 & 2314.660 & 0.44806 & - & 35 \\
SDSS J233325.92+152222.1 & 1.385 & 2500.000 & 0.50127& - & 25 \\
EX Hya &  1.637 & 4021.000 & 0.68231 & SXR & 1,2 21, 41 \\
\hline
\textit{Period Gap} & \textit{(2 h $<$ P$_{\rmn{orb}}$ $<$ 3 h)} \\
\hline
XSS J00564+4548 & 2.624 & 465.680 & 0.04929 & - & 7, 10 \\
\hline
\textit{Fast rotator} & \textit{(P$_{\rmn{spin}}$ / P$_{\rmn{orb}}$ $<$ 0.01)}\\
\hline
AE Aqr & 9.880 & 33.076 & 0.00093 & DP, Propeller & 11, 13, 26, 33, 41 \\
GK Per & 47.923 & 351.332 & 0.00204 & SXR, DP & 14, 21, 32, 33 \\
IGR J17303-0601 & 15.420 & 128.000 & 0.00231 & SXR & 2, 11 \\
DQ Her & 4.650 & 142.000 & 0.00848 & DP & 11, 33, 45 \\
XY Ari & 6.065 & 206.300 & 0.00945 & DP & 28, 33, 41 \\
\hline
\end{tabular}
\end{minipage}
\end{table*} 

\section{Conclusions}
We conclude that V2069~Cyg is an example of an IP that shows double-peaked emission profiles at the WD spin period which are probably caused by a weak magnetic field, in a WD with short spin period. The X-ray spectrum shows a soft bbody component and thermal plasma emission and its X-ray and optical emission have a double-peaked modulation. We performed simultaneous optical/X-ray observations of V2069~Cyg to search for any delays between these two energy bands. However the low count rates in the \textit{Swift} data did not allow to constrain these delays.

\section*{Acknowledgements}
Ilham Nasiroglu acknowledges support from the EU FP6 Transfer of Knowledge Project "Astrophysics of Neutron Stars" (MKTD-CT-2006-042722). Aga S\l{}owikowska acknowledges support from the grant N203 387737 of the Polish Ministry of Science and Higher Education, as well as the grant FNP HOM/2009/11B and the EU grant PERG05-GA-2009-249168. Gottfried Kanbach acknowledges support from the EU FP6 Transfer of Knowledge Project ASTROCENTER (MTKD-CT-2006-039965) and the kind hospitality of the Skinakas team at UoC. We acknowledge the use of public data from the Swift data archive. We would like to thank Aysun Akyuz and Arne Rau for discussions on this paper, as well as Anna Zajczyk (CAMK) and Andrzej Szary (UZG) for their help with observations. We thank the Skinakas Observatory for their support and allocation of telescope time. Skinakas Observatory is a collaborative project of the University of Crete, the Foundation for Research and Technology - Hellas, and the Max-Planck-Institute for Extraterrestrial Physics.


\begin{thebibliography}{99}
\bibitem[\protect\citeauthoryear{Allan et al.}{1996}]{Allan-etal1996} Allan A., Horne K., Hellier C., Mukai K., Barwig H., Bennie P.~J., Hilditch R.~W., 1996, MNRAS, 279, 1345 
\bibitem[\protect\citeauthoryear{Allan et al.}{1998}]{Allan-etal1998} Allan A., Hellier C., Beardmore A., 1998, MNRAS, 295, 167 
\bibitem[\protect\citeauthoryear{Anzolin et al.}{2008}]{Anzolin-etal2008} Anzolin G., de Martino D., Bonnet-Bidaud J.-M., et al., 2008, A\&A, 489, 1243
\bibitem[\protect\citeauthoryear{Arnaud}{1996}]{Arnaud1996} Arnaud K. A. 1996, ASPC, 101, 17 
\bibitem[\protect\citeauthoryear{Baskill et al.}{2005}]{Baskil-etal2005} Baskill D.~S., Wheatley P.~J., Osborne J.~P., 2005, MNRAS, 357, 626 
\bibitem[\protect\citeauthoryear{Burrows et al.}{2005}]{Burrows-etal2005} Burrows D. N., Hill J. E., Nousek J. A., et al. 2005a, Space Sci. Rev.,120, 165
\bibitem[\protect\citeauthoryear{Bonnet-Bidaud et al.}{2007}]{Bonnet-etal2007} Bonnet-Bidaud J.~M., de Martino D., Falanga M., Mouchet M., Masetti N., 2007, A\&A, 473, 185 
\bibitem[\protect\citeauthoryear{Bonnet-Bidaud et al.}{2009}]{Bonnet-etal2009} Bonnet-Bidaud J.~M., de Martino D., Mouchet M., 2009, ATel, 1895, 1 
\bibitem[\protect\citeauthoryear{Buckley et al.}{1995}]{Buckley-etal1995} Buckley D.~A.~H., Sekiguchi K., Motch C., et al., 1995, MNRAS, 275, 1028 
\bibitem[\protect\citeauthoryear{Burwitz et al.}{1996}]{Burwitz-etal1996} Burwitz V., Reinsch K., Beuermann K., Thomas H.-C., 1996, A\&A, 310, L25 
\bibitem[\protect\citeauthoryear{Butters et al.}{2007}]{Butters-etal2007} Butters O.~W., Barlow E.~J., Norton A.~J., Mukai K., 2007, A\&A, 475, L29 
\bibitem[\protect\citeauthoryear{Butters et al.}{2008}]{Butters-etal2008} Butters O.~W., Norton A.~J., Hakala P., Mukai K., Barlow E.~J., 2008, A\&A, 487, 271 
\bibitem[\protect\citeauthoryear{Butters et al.}{2009a}]{Butters-etal2009a} Butters O.~W., Katajainen S., Norton A.~J., Lehto H.~J., Piirola V., 2009a, A\&A, 496, 891 
\bibitem[\protect\citeauthoryear{Butters et al.}{2009b}]{Butters-etal2009b} Butters O.~W., Norton A.~J., Mukai K., Barlow E.~J., 2009b, A\&A, 498, L17 
\bibitem[\protect\citeauthoryear{Butters et al.}{2011}]{Butters-etal2011} Butters O.~W., Norton A.~J., Mukai K., Tomsick J.~A., 2011, A\&A, 526, A77 
\bibitem[\protect\citeauthoryear{Choi et al.}{1999}]{Choi-etal1999} Choi C.-S., Dotani T., Agrawal P.~C., 1999, ApJ, 525, 399 \bibitem[\protect\citeauthoryear{Crampton et al.}{1986}]{Crampton-etal1986} Crampton D., Fisher W.~A., Cowley A.~P., 1986, ApJ, 300, 788 
\bibitem[\protect\citeauthoryear{Cropper}{1990}]{Cropper1990} Cropper M., 1990, SSRv, 54, 195
\bibitem[\protect\citeauthoryear{de Martino et al.}{1999}]{deMartino-etal1999} de Martino D., Silvotti R., Buckley D.~A.~H., G{\"a}nsicke B.~T., Mouchet M., Mukai K., Rosen S.~R., 1999, A\&A, 350, 517
\bibitem[\protect\citeauthoryear{de Martino et al.}{2001}]{deMartino-etal2001} de Martino D., Matt G., Mukai K., et al. 2001, A\&A, 377, 499
\bibitem[\protect\citeauthoryear{de Martino et al.}{2006a}]{deMartino-etal2006a} de Martino D., Bonnet-Bidaud J.-M., Mouchet M., et al., 2006a, A\&A, 449, 1151 
\bibitem[\protect\citeauthoryear{de Martino et al.}{2006b}]{deMartino-etal2006b} de Martino D., Matt G., Mukai K., et al., 2006a, A\&A, 454, 287 
\bibitem[\protect\citeauthoryear{de Martino et al.}{2008}]{deMartino-etal2008} de Martino D., Matt G., Mukai K., et al., 2008, A\&A, 481, 149
\bibitem[\protect\citeauthoryear{de Martino et al.}{2009}]{deMartino-etal2009} de Martino D., Bonnet-Bidaud J.~M., Falanga M., Mouchet M., Motch C., 2009, ATel, 2089, 1 
\bibitem[\protect\citeauthoryear{Duck et al.}{1994}]{Duck-etal1994} Duck S.~R., Rosen S.~R., Ponman T.~J., Norton A.~J., Watson M.~G., Mason K.~O., 1994, MNRAS, 271, 372 
\bibitem[\protect\citeauthoryear{Dickey \& Lockman}{1990}]{Dickey-etal1990} Dickey J.~M., Lockman F.~J., 1990, ARA\&A, 28, 215 
\bibitem[\protect\citeauthoryear{Eracleous et al.}{1994}]{Eracleous-etal1994} Eracleous M., Horne K., Robinson E.~L., Zhang E.-H., Marsh T.~R., Wood J.~H., 1994, ApJ, 433, 313 
\bibitem[\protect\citeauthoryear{Evans \& Hellier}{2004}]{EvansHellier2004} Evans P.~A., Hellier C., 2004, MNRAS, 353, 447
\bibitem[\protect\citeauthoryear{Evans et al.}{2006}]{Evans-etal2006} Evans P.~A., Hellier C., Ramsay G., 2006, MNRAS, 369, 1229 
\bibitem[\protect\citeauthoryear{Evans \& Hellier}{2007}]{EvansHellier2007} Evans P.~A., Hellier C., 2007, ApJ, 663, 1277 
\bibitem[\protect\citeauthoryear{G{\"a}nsicke et al.}{2005}]{Gaensicke-etal2005} G{\"a}nsicke B. T., et al., 2005, MNRAS, 361, 141
\bibitem[\protect\citeauthoryear{Gregory \& Loredo}{1996}]{GregoryLoredo1996} Gregory P.~C., Loredo T.~J., 1996, ApJ, 473, 1059 
\bibitem[\protect\citeauthoryear{Haberl et al.}{2002}]{Haberl-etal2002} Haberl F., Motch C., Zickgraf F.-J., 2002, A\&A, 387, 201 
\bibitem[\protect\citeauthoryear{Harlaftis \& Horne}{1999}]{HarlaftisHorne1999} Harlaftis E.~T., Horne K., 1999, MNRAS, 305, 437 
\bibitem[\protect\citeauthoryear{Haswell et al.}{1997}]{Haswell-etal1997} Haswell C.~A., Patterson J., Thorstensen J.~R., Hellier C., Skillman D.~R., 1997, ApJ, 476, 847 
\bibitem[\protect\citeauthoryear{Hellier et al.}{1994}]{Hellier-etal1994} Hellier C., Ramseyer T.~F., Jablonski F.~J., 1994, MNRAS, 271, L25 
\bibitem[\protect\citeauthoryear{Hellier et al.}{1997}]{Hellier-etal1997} Hellier C., Mukai K., Beardmore A.~P., 1997, MNRAS, 292, 397 
\bibitem[\protect\citeauthoryear{Hellier, Beardmore, \& Buckley}{1998}]{Hellier-etal1998} Hellier C., Beardmore A.~P., Buckley D.~A.~H., 1998, MNRAS, 299, 851 
\bibitem[\protect\citeauthoryear{Hellier}{2001}]{Hellier2001} Hellier C., 2001, Cataclysmic Variable Stars, Springer-Praxis, Chichester, UK
\bibitem[\protect\citeauthoryear{Hellier}{2007}]{Hellier2007} Hellier C., 2007, IAU Symp. 243, Star–Disk Interaction in Young Stars, p. 325, Keele University, UK
\bibitem[\protect\citeauthoryear{Hellier et al.}{2002a}]{Hellier-etal2002a} Hellier C., Wynn G.~A., Buckley D.~A.~H., 2002a, MNRAS, 333, 84
\bibitem[\protect\citeauthoryear{Hellier et al.}{2002b}]{Hellier-etal2002b} Hellier C., Beardmore A.~P., Mukai K., 2002b, A\&A, 389, 904
\bibitem[\protect\citeauthoryear{Hilton et al.}{2009}]{Hilton-etal2009} Hilton E.~J., Szkody P., Mukadam A., Henden A., Dillon W., Schmidt G.~D., 2009, AJ, 137, 3606 
\bibitem[\protect\citeauthoryear{Israel et al.}{2003}]{Israel-etal2003} Israel G. L., et al., 2003, ApJ, 598, 492 
\bibitem[\protect\citeauthoryear{Kanbach et al.}{2003}]{Kanbach-etal2003} Kanbach G., Kellner S., Schrey F.~Z., Steinle H., Straubmeier C., Spruit H.~C., 2003, SPIE, 4841, 82 
\bibitem[\protect\citeauthoryear{Kemp et al.}{2002}]{Kemp-etal2002} Kemp J., Patterson J., Thorstensen J.~R., Fried R.~E., Skillman D.~R., Billings G., 2002, PASP, 114, 623 
\bibitem[\protect\citeauthoryear{Kim et al.}{2005}]{Kim-etal2005} Kim Y.~G., Andronov I.~L., Park S.~S., Jeon Y.-B., 2005, A\&A, 441, 663
\bibitem[\protect\citeauthoryear{Mauche}{2004}]{Mauche2004} Mauche C. W., 2004, in Vrielmann S., Cropper M., eds, ASP Conf. Ser. Vol. 315, Magnetic Cataclysmic Variables. Astron. Soc. Pac., San Francisco, p. 120
\bibitem[\protect\citeauthoryear{Mewe et al. 1985}{}]{Mewe-etal1985} Mewe R., Gronenschild E.~H.~B.~M., van den Oord G.~H.~J., 1985, A\&AS, 62, 197 
\bibitem[\protect\citeauthoryear{Motch et al.}{1996}]{Motch-etal1996} Motch C., Haberl F., Guillout P., Pakull M., Reinsch K., Krautter J., 1996, A\&A, 307, 459 
\bibitem[\protect\citeauthoryear{Norton et al.}{1999}]{Norton-etal1999} Norton A.~J., Beardmore A.~P., Allan A., Hellier C., 1999, A\&A, 347, 203 
\bibitem[\protect\citeauthoryear{Norton et al.}{2002}]{Norton-etal2002} Norton A.~J., Quaintrell H., Katajainen S., Lehto H.~J., Mukai K., Negueruela I., 2002, A\&A, 384, 195 
\bibitem[\protect\citeauthoryear{Norton et al.}{2004a}]{Norton-etal2004a} Norton A.~J., Somerscales R.~V., Parker T.~L., Wynn G.~A., West R.~G., 2004a, RMxAC, 20, 138
\bibitem[\protect\citeauthoryear{Norton et al.}{2004b}]{Norton-etal2004b} Norton A.~J., Haswell C.~A., Wynn G.~A., 2004b, A\&A, 419, 1025 
\bibitem[\protect\citeauthoryear{Norton et al.}{2004c}]{Norton-etal2004c} A. J., Somerscales R. V., Wynn G. A., 2004c, in Vrielmann S., Cropper M., eds, ASP Conf. Ser. Vol. 315, Magnetic Cataclysmic Variables. Astron. Soc. Pac., San Francisco, p. 216 
\bibitem[\protect\citeauthoryear{Norton \& Mukai}{2007}]{NortonMukai2007} Norton A.~J., Mukai K., 2007, A\&A, 472, 225 
\bibitem[\protect\citeauthoryear{Parker et al.}{2005}]{Parker-etal2005} Parker T.~L., Norton A.~J., Mukai K., 2005, A\&A, 439, 213
\bibitem[\protect\citeauthoryear{Patterson}{1994}]{Patterson1994} Patterson J., 1994, PASP, 106, 209 
\bibitem[\protect\citeauthoryear{Patterson et al.}{2004}]{Patterson-etal2004} Patterson J., et al., 2004, PASP, 116, 516 
\bibitem[\protect\citeauthoryear{Patterson et al.}{2011}]{Patterson-etal2011} Patterson J., et al., 2011, PASP, 123, 130
\bibitem[\protect\citeauthoryear{Pretorius}{2009}]{Pretorius2009} Pretorius M.~L., 2009, MNRAS, 395, 386 
\bibitem[\protect\citeauthoryear{Revnivtsev et al.}{2010}]{Revnivtsev-etal2010} Revnivtsev M., et al., 2010, A\&A, 513, A63 
\bibitem[\protect\citeauthoryear{Rosen et al.}{1988}]{Rosen-etal1988} Rosen S.~R., Mason K.~O., Cordova F.~A., 1988, MNRAS, 231, 549
\bibitem[\protect\citeauthoryear{Scaringi et al.}{2010}]{Scaringi-etal2010} Scaringi S., et al., 2010, MNRAS, 401, 2207
\bibitem[\protect\citeauthoryear{Scaringi et al.}{2011}]{Scaringi-etal2011} Scaringi S., et al., 2011, A\&A, 530, A6 
\bibitem[\protect\citeauthoryear{Schlegel}{2005}]{Schlegel2005} Schlegel E.~M., 2005, A\&A, 433, 635 
\bibitem[\protect\citeauthoryear{Staude et al.}{2003}]{Staude-etal2003} Staude A., Schwope A.~D., Krumpe M., Hambaryan V., Schwarz R., 2003, A\&A, 406, 253 
\bibitem[\protect\citeauthoryear{Staude et al.}{2008}]{Staude-etal2008} Staude A., Schwope A.~D., Schwarz R., Vogel J., Krumpe M., Nebot Gomez-Moran A., 2008, A\&A, 486, 899
\bibitem[\protect\citeauthoryear{Str{\"u}der etal.}{2001}]{Strueder-etal2001} Str{\"u}der L., et al., 2001, A\&A, 365, L18
\bibitem[\protect\citeauthoryear{Thorstensen \& Taylor}{2001}]{ThorstensenTaylor2001} Thorstensen J.~R., Taylor C.~J., 2001, MNRAS, 326, 1235 
\bibitem[\protect\citeauthoryear{Thorstensen, Peters,\& Skinner}{2010}]{Thorstensen-etal2010} Thorstensen J.~R., Peters C.~S., Skinner J.~N., 2010, PASP, 122, 1285 
\bibitem[\protect\citeauthoryear{Turner et al.}{2001}]{Turner-etal2001} Turner M. J. L., et al., 2001, A\&A, 365, L27 
\bibitem[\protect\citeauthoryear{Warner}{1995}]{Warner1995} Warner B., 1995, Cataclysmic Variable Stars, Cambridge Univ. Press, Cambridge, UK
\bibitem[\protect\citeauthoryear{Zhang et al.}{1995}]{Zhang-etal1995} Zhang E., Robinson E.~L., Stiening R.~F., Horne K., 1995, ApJ, 454, 447
\end{thebibliography}
\end{document}